\documentclass[aps,prb,twocolumn,amsmath,amssymb,superscriptaddress]{revtex4}

\usepackage{graphicx}
\usepackage{dcolumn}
\usepackage{bm}
\usepackage{color}
\usepackage{subfigure}
\usepackage{amsmath}
\usepackage{amssymb}
\usepackage{setspace}
\newcommand{\ket}[1]{|#1\rangle}                      
\newcommand{\bra}[1]{\langle #1|}                     
\begin{document}
\title{Quantum walks and wavepacket dynamics on a lattice with twisted photons}
\author{Filippo Cardano}
\affiliation{Dipartimento di Fisica, Universit\`{a} di Napoli Federico II, Complesso Universitario di Monte Sant'Angelo, Napoli, Italy}
\author{Francesco Massa}
\altaffiliation[Current address: ]{Faculty of Physics, University of Vienna, Boltzmanngasse 5, 1090, Vienna, Austria}
\affiliation{Dipartimento di Fisica, Universit\`{a} di Napoli Federico II, Complesso Universitario di Monte Sant'Angelo, Napoli, Italy}
\author{Hammam Qassim}
\affiliation{Department of Physics, University of Ottawa, 150 Louis Pasteur, Ottawa, Ontario, K1N 6N5 Canada}
\author{Ebrahim Karimi}
\affiliation{Department of Physics, University of Ottawa, 150 Louis Pasteur, Ottawa, Ontario, K1N 6N5 Canada}
\author{Sergei Slussarenko}
\altaffiliation[Current address: ]{Centre for Quantum Dynamics, Griffith University, Brisbane 4111, Australia}
\affiliation{Dipartimento di Fisica, Universit\`{a} di Napoli Federico II, Complesso Universitario di Monte Sant'Angelo, Napoli, Italy}
\author{Domenico Paparo}
\affiliation{CNR-SPIN, Complesso Universitario di Monte Sant'Angelo, Napoli, Italy}
\author{Corrado de Lisio}
\affiliation{Dipartimento di Fisica, Universit\`{a} di Napoli Federico II, Complesso Universitario di Monte Sant'Angelo, Napoli, Italy}
\affiliation{CNR-SPIN, Complesso Universitario di Monte Sant'Angelo, Napoli, Italy}
\author{Fabio Sciarrino}
\affiliation{Dipartimento di Fisica, Sapienza Universit\`{a} di Roma, Roma 00185, Italy}
\author{Enrico Santamato}
\affiliation{Dipartimento di Fisica, Universit\`{a} di Napoli Federico II, Complesso Universitario di Monte Sant'Angelo, Napoli, Italy}
\author{Robert W. Boyd}
\affiliation{Department of Physics, University of Ottawa, 150 Louis Pasteur, Ottawa, Ontario, K1N 6N5 Canada}
\author{Lorenzo Marrucci}
\affiliation{Dipartimento di Fisica, Universit\`{a} di Napoli Federico II, Complesso Universitario di Monte Sant'Angelo, Napoli, Italy}
\affiliation{CNR-SPIN, Complesso Universitario di Monte Sant'Angelo, Napoli, Italy}
\begin{abstract}
The ``quantum walk'' has emerged recently as a paradigmatic process for the dynamic simulation of complex quantum systems, entanglement production and quantum computation. Hitherto, photonic implementations of quantum walks have mainly been based on multi-path interferometric schemes in real space. Here, we report the experimental realization of a discrete quantum walk taking place in the orbital angular momentum space of light, both for a single photon and for two simultaneous photons. In contrast to previous implementations, the whole process develops in a single light beam, with no need of interferometers; it requires optical resources scaling linearly with the number of steps; and it allows flexible control of input and output superposition states. Exploiting the latter property, we explored the system band structure in momentum space and the associated spin-orbit topological features by simulating the quantum dynamics of Gaussian wavepackets. Our demonstration introduces a novel versatile photonic platform for quantum simulations.
\end{abstract}
\maketitle

\section*{Introduction}
First proposed by Feynman about thirty years ago\cite{Feyn82}, the simulation of a complex quantum system by means of another simpler and well controlled quantum system is nowadays becoming a feasible, although still challenging task. Photons are a reliable resource in this arena, as witnessed by the large variety of photonic architectures that have been introduced hitherto for the realization of quantum simulators\cite{Walt12_NatPhys}. Among simulated processes, the quantum walk\cite{Kemp03_ConPhys} (QW) is receiving a wide interest. A QW can be interpreted as the quantum counterpart of the well known classical random walk. In its simplest, discrete and one-dimensional (1D) example, the latter is a path consisting of a sequence of random steps along a line. At each step, the walker moves forward or backward according to the outcome of a random process, such as the flip of a coin. When both the walker and the coin are quantum systems we obtain a QW. The final probability distribution for the walker position shows striking differences with respect to the classical process, due to interferences between coherent superpositions of different paths\cite{Knig06_PRA}. It has been demonstrated that this quantum process can be used to perform quantum search algorithms on a graph\cite{Shen03_PRA,Poto09_PRA} and universal quantum computation\cite{Chil09_PRL,Love10_PRA}. Moreover, it represents a versatile approach to the simulation of phenomena characterizing complex systems, such as Anderson localization in disordered media \cite{Cres13_NatPhot} and energy transport in chemical processes \cite{Mohs08_JCP}. The coin-walker interaction, for example, gives rise to fascinating analogies with quantum effects arising from spin-orbit coupling: recently, it was demonstrated that discrete QWs can simulate all classes of topological phases in 1D and 2D \cite{Kita10_PRA} and topologically-protected bound states have been observed at the interface between regions with different topologies\cite{Kita12_NatCom}.

In the last decade, implementations of QWs in 1D have been realized in a variety of physical systems, such as trapped ions\cite{Schm09_PRL,Zahr10_PRL} or atoms\cite{Kars09_Sci}, nuclear-magnetic resonance (NMR) systems\cite{Ryan05_PRA}, and photons, using both bulk optics\cite{Zhan07_PRA,Broo10_PRL,Schr10_PRL} and integrated waveguides\cite{Peru10_Sci,Owen11_NJP,Sans12_PRL}. Remarkably, only a few photonic simulations of multi-particles QWs have been reported, using two-photon states\cite{Peru10_Sci,Owen11_NJP,Sans12_PRL,Cres13_NatPhot} or classical coherent sources\cite{Schr12_Sci}. In photonic architectures different strategies can be adopted, according to the optical degrees of freedom exploited to encode the coin and the walker quantum systems. In 2010 Zhang \textit{et al.} proposed a novel approach for the realization of a photonic QW, based on the idea of encoding the coin and the walker in the spin angular momentum (SAM) and in the orbital angular momentum (OAM) of light, respectively\cite{Zhan10_PRA}. A possible implementation of the same idea in a loop-based configuration has been also analyzed\cite{Goya13_PRL}. These theoretical proposals put forward for the first time the possibility of implementing a photonic walk without interferometers, with the whole process taking place within a single light beam (we refer here to ``real-space interferometers'', i.e.\ relying on optical path splitting, as any kind of wave propagation involves some form of modal interference). To obtain this result, these schemes rely on the spin-orbit coupling occurring in a special optical element called q-plate\cite{Marr06_PRL}, whose action will be discussed later on. In the present work, we implement experimentally the proposal by Zhang \textit{et al.}, thus demonstrating the first photonic QW occurring in a single light beam and using the OAM degree of freedom of photons as discrete walker coordinate (we notice that, although the QW realized in Refs.\ \onlinecite{Schr10_PRL,Schr12_Sci} involves only inner degrees of freedom of a single light beam, its actual implementation still relies on splitting the beam in a spatial interferometer). We demonstrate both the QW of single photons and that of two indistinguishable photons, thus highlighting the role of multiparticle quantum interferences. As we will discuss further below, this novel implementation has potential advantages in terms of stability and scalability. Moreover, in contrast to most current integrated-optics approaches, it allows one to vary dynamically the system Hamiltonian and to measure the whole evolution step by step (not only the final output), without changing the experimental setup. Finally, a very important feature of this QW implementation is the possibility of flexibly preparing arbitrary superpositions or ``delocalized'' initial states of the walker, by exploiting standard holographic optical devices (or conversely to make a full quantum tomography of the delocalized output quantum state). As a specific demonstration of this feature, we experimentally verified the band structure characterizing a QW; we prepared  Gaussian wavepackets of a photon in OAM space, for different values of the average linear quasi-momentum, and observed their free quantum dynamics, governed by the underlying band dispersion relations and the associated topological spin-orbit features\cite{Abal06_PRA,Kita10_PRA,Kita12_QIP}.

\section*{Results}
\subsection*{Quantum walk in the OAM space of a photon}
In the quantum theory framework, a discrete QW typically involves a system described by a Hilbert space $\mathcal H$ obtained by the direct product $\mathcal H_c\otimes\mathcal H_w$ of the coin and the walker subspaces, respectively. In the simplest case, the walker is moving in a 1D lattice and, at each step, has only two choices. Accordingly, the subspace $\mathcal H_c$ is two-dimensional (2D), while $\mathcal H_w$ is infinite-dimensional; they are spanned by the vectors $\{\ket{\uparrow}_c,\ket{\downarrow}_c\}$ and $\{\ket{x}_w,\; x\in \mathbb{Z}\}$, respectively (in the following, subscripts $c$ and $w$ will be omitted for brevity whenever there is no risk of ambiguity). Alternatively, the walker state can be described in terms of its quasi-momentum $k$, which is defined in the first Brillouin zone $k\in(-\pi,\pi)$. The relation between the two representations is given by the discrete Fourier transform, i.e. $\ket{k}=(1/\sqrt{2\pi})\sum_x{e^{-ikx}\ket{x}}$. The momentum representation provides the framework to analyze the effective band structure of the QW system, as will be discussed later on. The displacement of the walker at each step of the process is realized by the shift operator $\hat S$
\begin{equation}\label{main:shift}
\hat S=\ket{\uparrow}\bra{\uparrow}\otimes\hat L^++\ket{\downarrow}\bra{\downarrow}\otimes\hat L^-,
\end{equation} 
where the operators $\hat L^{\pm}$ shift the position of the walker, i.e.\ $\hat L^{\pm}\ket{x}=\ket{x\pm1}$. The displacement introduced by $\hat S$ is conditioned by the coin; when this is in the state $\ket{\uparrow}$, the walker moves up, or vice versa. As a consequence, the operator $\hat S$ entangles the coin and the walker systems\cite{Abal06_PRA,Viei13_PRL}. Between consecutive displacements, the ``randomness'' is introduced by a unitary operator $\hat T$ acting on the coin subspace, as generally given by $\hat T\,\ket{\uparrow}=a\ket{\uparrow}+b\ket{\downarrow}$ and 
$\hat T\,\ket{\downarrow}=b^*\ket{\uparrow}-a^*\ket{\downarrow}$ (up to a global phase), with $a,b$ complex numbers such that $|a|^2+|b|^2=1$; for an unbiased walk $|a|=|b|=1/\sqrt{2}$. A single step of the walk is described by the step operator $\hat U=\hat S\cdot(\hat T\otimes \hat I_w)$, where $\hat I_w$ is the identity operator in $\mathcal H_w$. After $n$ steps, the system initially prepared in the state $\ket{\psi_0}$ evolves to a new state
\begin{equation}
\ket{\psi_n}= \hat U^n\ket{\psi_0}.
\end{equation} 
In the momentum representation, eigenstates of the operator $\hat U$ have a simple expression\cite{Abal06_PRA,Kita10_PRA}, given by $\ket{k,s}=\ket{\phi_s(k)}_c\otimes\ket{k}_w$ with eigenvalues $e^{-i\omega_s(k)}$, where $s\in\{1,2\}$. The dispersion relation of quasi-energies $\omega_s(k)$ shows two gapped bands associated to the coin eigenstates $\ket{\phi_s(k)}$, while the parametric dependence of the latter states on $k$ defines the QW topological structure\cite{Kita12_QIP}. A more detailed analysis of these QW properties is provided in the Supplementary Materials (SM).

\begin{figure}[t]
\centering
\includegraphics[width=8.5cm]{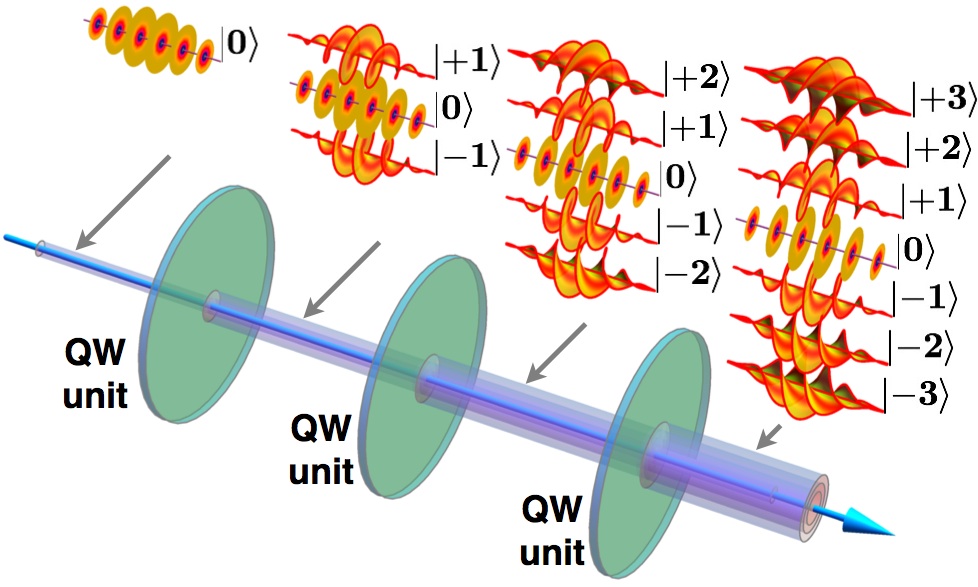}
\caption{Conceptual scheme of the single-beam photonic quantum walk in the space of OAM. In each traversed optical stage (QW unit), the photon can move to an OAM value $m$ that is increased or decreased by one unit (or stay still, in the hybrid configuration). The OAM decomposition of the photonic wavefunction at each stage thus includes many different components, as shown in the call-outs in which modes having different OAM values are represented by the corresponding helical (or ``twisted'') wavefronts.}
\label{fig:naive}
\end{figure}

Consider now a photon and its internal degrees of freedom represented by the SAM and the OAM. In the limit of paraxial optics, these two quantities are independent and well defined; the first is associated with the polarization of the light, while the second is related to the azimuthal structure of the optical wavefront in the transverse plane \cite{Marr11_JOpt}. The SAM space is spanned by vectors $\{\ket{L},\ket{R}\}$, representing left-circular and right-circular polarizations. The OAM space is spanned by vectors $\ket{m}$ with $m\in \mathbb{Z}$, which denote a photon carrying $m\hbar$ of OAM along the propagation axis, where $\hbar$ is the reduced Planck constant, and having a correspondingly ``twisted'' wavefunction (see Fig.\ \ref{fig:naive}).

%
%
\begin{figure}[t]
\centering
\includegraphics[width=8.5cm]{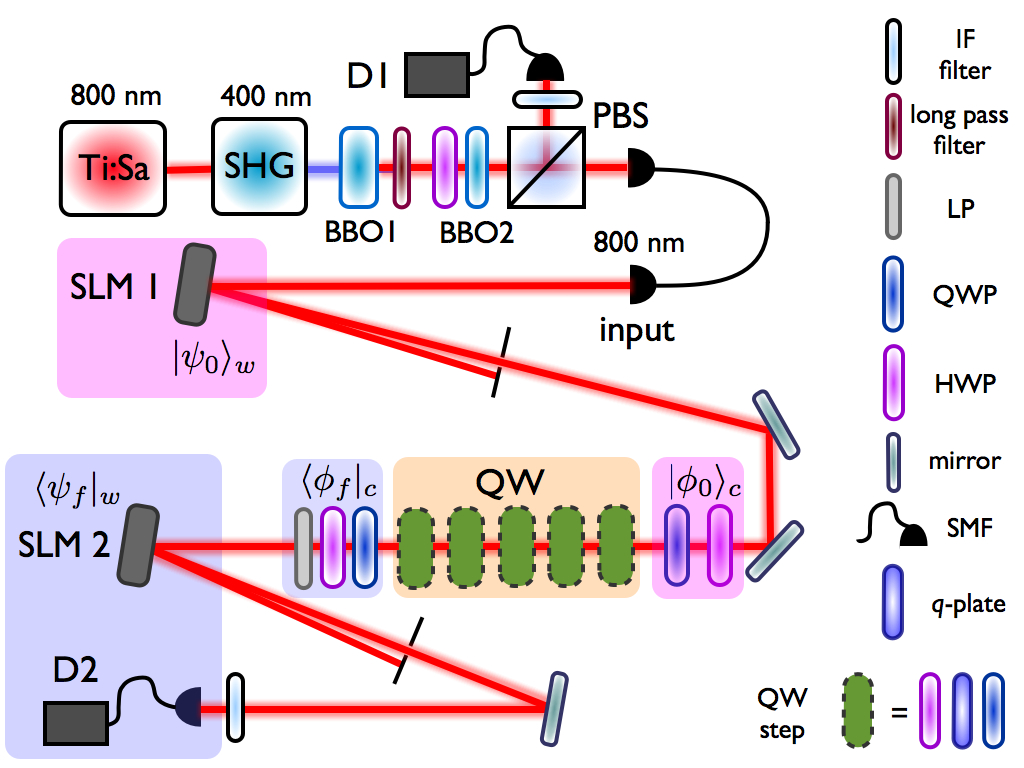}
\caption{Experimental apparatus for single-photon QW experiments. Frequency-doubled laser pulses at 400 nm and with 140 mW average power, obtained from the fundamental pulses (100 fs) generated by a Titanium:Sapphire source (Ti:Sa) at a repetition rate of 82 MHz, pump a 3-mm-thick nonlinear $\beta$-barium borate crystal (BBO1) cut for type-II SPDC.  Photon pairs at 800 nm generated through this process, cleaned from residual radiation at 400 nm using a long pass filter, pass through a HWP and the BBO2 crystal (cut as BBO1, but 1.5-mm-thick) in order to compensate both spatial and temporal walk-off introduced by BBO1. Next, the two photons are split by a PBS; one is sent directly to the avalanche single-photon detector (APD) D1, while the other is coupled into a SMF. At the exit of the fiber, the photon goes through $N$ identical subsequent QW steps ($N=5$ in the figure), is then analyzed in both polarization and OAM and is finally detected with APD D2, in coincidence with D1. Before entering the first QW step, a spatial light modulator (SLM 1) and a HWP-QWP set are used to prepare the photon initial state in the OAM and SAM spaces, respectively. At the exit of the last step, the polarization projection on the state $\ket{\phi_f}_c$ is performed with a second HWP-QWP set followed by a linear polarizer (LP). The OAM state is then analyzed by diffraction on SLM 2, followed by coupling into a SMF. The projection state $\ket{\psi_f}_w$ corresponding to each OAM eigenvalue $m$ was thus fixed by the hologram pattern displayed on SLM 2. Before detection, interferential filters (IF) centered at 800 nm and with a bandwidth of 3.6 nm were used for spectral cleaning. As shown in the legend, a single QW step consists of a QWP (optical axis at 45$^{\circ}$ from the horizontal), a q-plate with $q=1/2$ (axis at 0$^{\circ}$), and a HWP (axis at 0$^{\circ}$); the HWP was not included in the wavepacket and two-photon experiments.}
\label{fig:layout}
\end{figure}
In our implementation, the coin and the walker systems are encoded in the SAM and the OAM of a photon, respectively. In particular, the spatial walker coordinate $x$ is replaced by the OAM coordinate $m$. The concept of a QW in OAM within a single optical beam is pictorially illustrated in Fig.\ \ref{fig:naive}. The step operator $\hat U$ is realized by means of linear-optical elements. In the coin subspace, the unitary operator $\hat{T}$ can be implemented by birefringent plates, such as quarter-wave plates (QWP) and/or half-wave plates (HWP). In particular, we used only waveplate combinations giving rise to unbiased QWs. The shift operator $\hat S$ is realized by a q-plate (QP), a recently-introduced photonic device which has already found many useful applications in classical and quantum optics \cite{Marr06_PRL,Marr11_JOpt,Card12_AO,Damb12_NatComm,Damb13_NatComm}. The QP is a birefringent liquid-crystal medium with an inhomogeneous optical axis that has been arranged in a singular pattern, with topological charge $q$, so as to give rise to an engineered spin-orbit coupling in the light crossing it. In particular, the QP raises or lowers the OAM of the incoming photon according to its SAM state, while leaving the photon in the same optical beam, i.e.\ with no deflections nor diffractions. In the actual device, the radial profile of the photonic wave function undergoes a small alteration, which however can be approximately neglected in our implementation, as discussed in the SM. More precisely, the action of a QP can be generally described by the operator $\hat Q_\delta$
\begin{eqnarray}\label{main:qplate}
\hat Q_\delta \ket{L,m} &= \cos{(\delta/2)}\ket{L,m}-i\sin{(\delta/2)}\ket{R,m+2q} \nonumber\\
\hat Q_\delta\ket{R,m}&= \cos{(\delta/2)}\ket{R,m}-i\sin{(\delta/2)}\ket{L,m-2q}, \label{exp:qplate}
\end{eqnarray}
where $q$ is the topological charge of the QP and $\delta$ is the optical birefringent phase-retardation \cite{Marr06_PRL,Marr11_JOpt}. Whereas $q$ is a fixed property of the q-plate, $\delta$ can be controlled dynamically by tuning an applied voltage \cite{Picc10_APL}. As shown in Eq.\ \ref{main:qplate}, the action of the q-plate is made of two terms. The first, proportional to $\cos(\delta/2)$, leaves the photon in its input state. The second, proportional to $\sin(\delta/2)$, implements the conditional displacement of Eq.\ \ref{main:shift}, but also adds a flip of the coin state. The latter effect can be compensated by inserting an additional HWP. When $\delta=\pi$ (``standard'' configuration) the first term vanishes and the standard shift operator $\hat S$ is obtained. When $\delta=0$, the evolution is trivial (the walker stands still), while for intermediate values $0<\delta<\pi$ we have a novel kind of evolution: besides moving forward or backward, the walker at each step is provided with a third option, that is to remain in the same position. We refer to this as a ``hybrid'' configuration, since it mimics a walk with three possible choices, although the coin is still two-dimensional. Similar to an effective mass, the $\delta$ parameter controls the degree of mobility of the walker, ranging from a vanishing mobility for $\delta=0$ to a maximal mobility (not taking into account the effect of the coin) for $\delta=\pi$.
%
%
\begin{figure*}[thb]
\centering
\includegraphics[width=15cm]{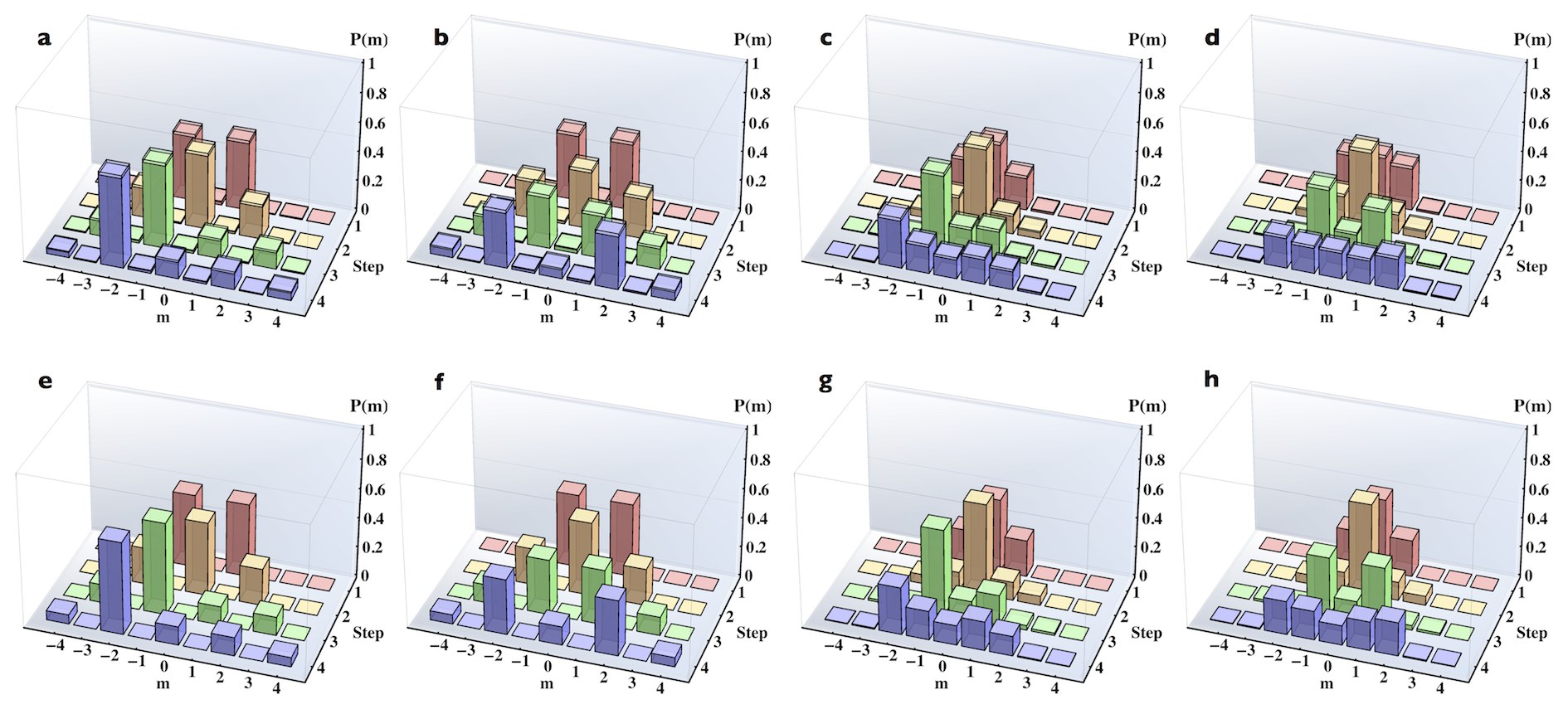}
\caption{Four-step quantum walk for a single photon with localized input. a-d) Experimental results, including both intermediate and final probabilities for different OAM states in the evolution (summed over different polarizations). The intermediate probabilities at step $n$ are obtained by switching off all QPs that follow that step, that is setting $\delta=0$. Panels a) and b) refer to the standard case with two different input states for the coin subsystem, $(\alpha,\beta)= (0,1)$ and $1/\sqrt{2} (1,i)$, respectively. c) and d) refer to the hybrid case with $\delta=1.57$, with the same initial coin-states. e-h) Corresponding theoretical predictions. Poissonian statistical uncertainties at plus-or-minus one standard deviation are shown as transparent-volumes in panels a-e. The similarities between experimental and predicted final OAM distributions are $(94.7\pm0.4)\%$, $(93.4\pm0.5)\%$, $(99.7\pm0.1)\%$ and $(99.2\pm0.2)\%$, respectively. Panels on the same column refer to the same configuration and initial states. The color scale reflects the number of steps.}
\label{fig:data1}
\end{figure*}
The photon entering the QW setup is initially prepared in a separable state $\ket{\psi_0}=\ket{\phi_0}_c\otimes\ket{\psi_0}_w$. A computer-generated hologram shown on a spatial light modulator (SLM 1) is used to prepare the walker initial state in a generic superposition of OAM states\cite{Bold13_OL,Damb13_SciRep} in $H_o$ (see the SM for details). After the SLM 1, the coin is prepared in the state $\ket{\phi_0}_c=\alpha\ket{L}+\beta\ket{R}$, where the two complex coefficients $\alpha$ and $\beta$ (with $|\alpha|^2+|\beta|^2=1$) can be selected at will by a QWP-HWP set (apart from an unimportant global phase). The photon then undergoes the QW evolution and, at the exit, is analyzed in both polarization and OAM so as to determine the output probabilities. Details on projective measurements in OAM are given in SM.

\subsection*{Single-photon quantum walk with localized initial state}
In our first experiment, the step operator $\hat U$ is implemented by a sequence of a QWP, a QP, and a HWP. The QPs have $q=1/2$, so as to induce OAM shifts of $\pm1$. Due to reflection losses (mainly at the QP, which is not antireflection-coated), each step has a transmission efficiency of 86\% (but adding an antireflection coating could easily improve this value to $>95$\%). The $n$-step walk is then implemented by simply cascading a sequence of QWP-QP-HWP on the single optical axis of the system. In the implemented setup, the linear distance $d$ between adjacent steps is small compared to the Rayleigh range $z_R$ of the photons, i.e.\ $d/z_R\ll1$ (near-field regime), so as to avoid optical effects that would alter the nature of the simulated process; a detailed discussion is provided in the SM. The layout of the apparatus is shown in Fig.\ \ref{fig:layout}. A photon pair is generated by spontaneous parametric down-conversion (SPDC) in the product state $\ket{H}\ket{V}$, where $H$ and $V$ stand for horizontal and vertical linear polarization (see the caption of Fig.\ \ref{fig:layout} for details). To carry out a single-particle QW simulation, we split the two input photons with a polarizing beam splitter (PBS); the $H$-polarized photon only enter the QW setup, after being coupled into a single-mode optical fiber (SMF), which sets $m=0$. At the exit of the fiber, the initial polarization of the photon is recovered using a QWP-HWP set (not shown in the figure). The $V$-polarized photon, reflected at the PBS, is sent directly to a detector and provides a trigger, so as to operate the QW simulation in a heralded single-photon quantum regime.

As a first set of experiments, we carried out QWs with single photons prepared in the localized state $m=0$ on the OAM lattice, with varying SAM input states. In Fig.\ \ref{fig:data1} we report the experimental and predicted results relative to a 4-step QW, for two possible input polarization states, and both in the standard and hybrid configurations (two additional input polarization cases are given in Fig.\ \ref{fig:dataSM}in the SM). To evaluate quantitatively the agreement between measured and predicted probability distributions, $P(m)$ and $P'(m)$, we also computed their ``similarity''\\
$S=\left(\sum_{m}\sqrt{P(m)P'(m)}\right)^2 / \left(\sum_{m}P(m)\sum_{m}P'(m)\right)$. The values we obtain in the various cases are given in the figure captions.

\subsection*{Simulation of wavepacket dynamics in OAM space}
Next, exploiting the possibility to control the walker initial state, we investigated the QW evolution for states with a given quasi-momentum $k$, thus probing the dispersion relation of the effective band structure of our QW system and its associated topological structure (for the standard case with $\delta=\pi$). A similar approach was used to simulate the evolution of a multi-band Bloch particle in a time-dependent field, by shining an engineered wave-guide array with classical coherent light \cite{Long06_PRB} (see also Ref.\ \onlinecite{Chri03_Nat} for a  review on discrete-waveguide lattice effects). Controlling the quasi-momentum of delocalized quantum states is crucial for carrying out quantum simulations of Bloch-particle dynamics, as shown for instance in Ref.\ \onlinecite{Atal13_NatPhys}.

Using the holographic method described in the SM, we prepared single-photon wavepackets given by $\ket{\psi_0^s}=\ket{\phi_s(k_0)}_c \otimes \left(\sum_m {A(m)\,e^{-ik_0 m} \ket{m}_w}\right)$, where $A(m)=A_0 e^{-m^2/2\sigma^2}$ is a Gaussian envelope in OAM space and $\ket{\phi_s(k_0)}_c$ (with $s=1,2$) is the polarization-coin part of the step-operator eigenstate\cite{Kita10_PRA}. The associated quasi-momentum has a Gaussian distribution centered on $k_0$ (incidentally, the average quasi-momentum $k_0$ corresponds also to the average azimuthal angle in real space for the optical field distribution within the beam). When $A(m)$ is a slowly varying envelope, these wavepackets are expected to propagate on the 1D lattice with only minimal shape variations and with a speed given by the group velocity $V_s(k_0)=(d\omega_s/d k)_{k=k_0}$. Interestingly states belonging to different bands, which correspond to orthogonal polarization eigenstates, propagate in opposite directions, i.e. $V_1(k_0)=-V_2(k_0)$, highlighting the strong spin-orbit coupling of this system (see SM for more details).
%
%
\begin{figure}[t]
\centering
\includegraphics[width=8.5cm]{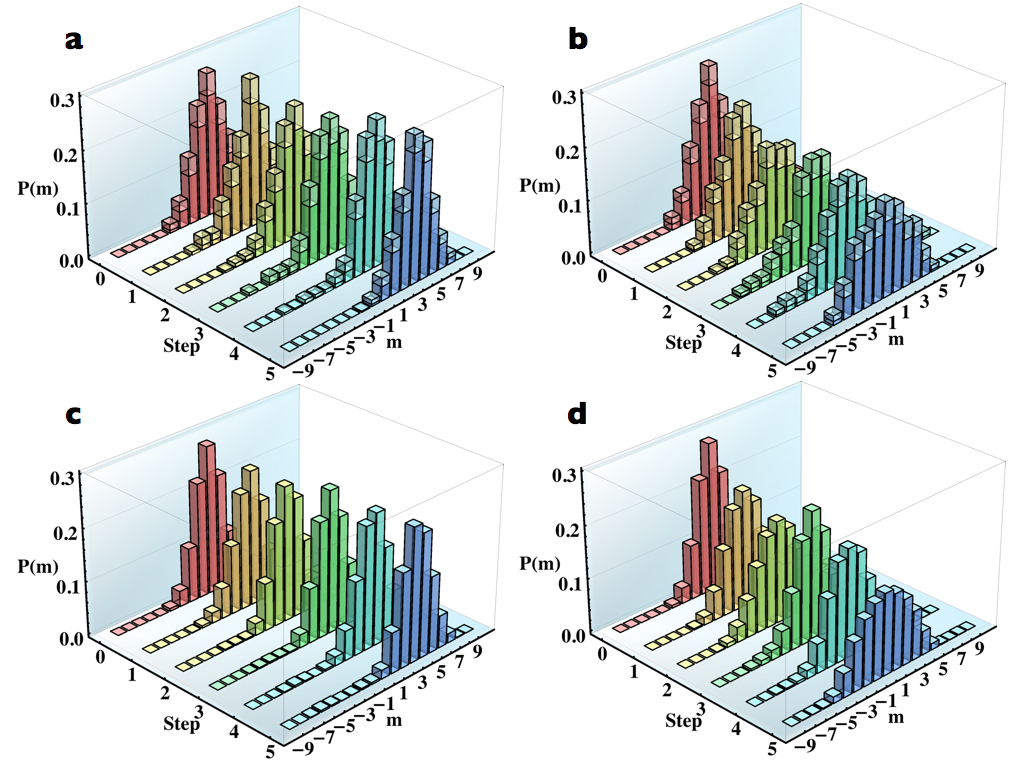}
\caption{Wavepacket propagation in a five-steps quantum walk. a-b) Experimental results, showing the step-by-step evolution of the OAM distribution of a single photon prepared in a Gaussian wavepacket with $\sigma=2$, in the SAM band $s=1$ (summed over different polarizations). Panels a) and b) correspond to the two cases $k_0=\pi$ (maximal group velocity) and $k_0=\pi/2$ (vanishing group velocity), respectively. The latter configuration shows some spreading of the Gaussian envelope, governed by the group-velocity dispersion. Poissonian statistical uncertainties at plus-or-minus one standard deviation are shown as transparent-volumes. c-d) Theoretical predictions corresponding to the same cases. At the fifth step, the similarity between experimental and theoretical OAM distributions are  $(98.2\pm0.4)\%$ and $(99.0\pm0.2)\%$, respectively. The color scale reflects the number of steps.} \label{fig:packets}
\end{figure}
\begin{figure*}[thb]
\centering
\includegraphics[width=14cm]{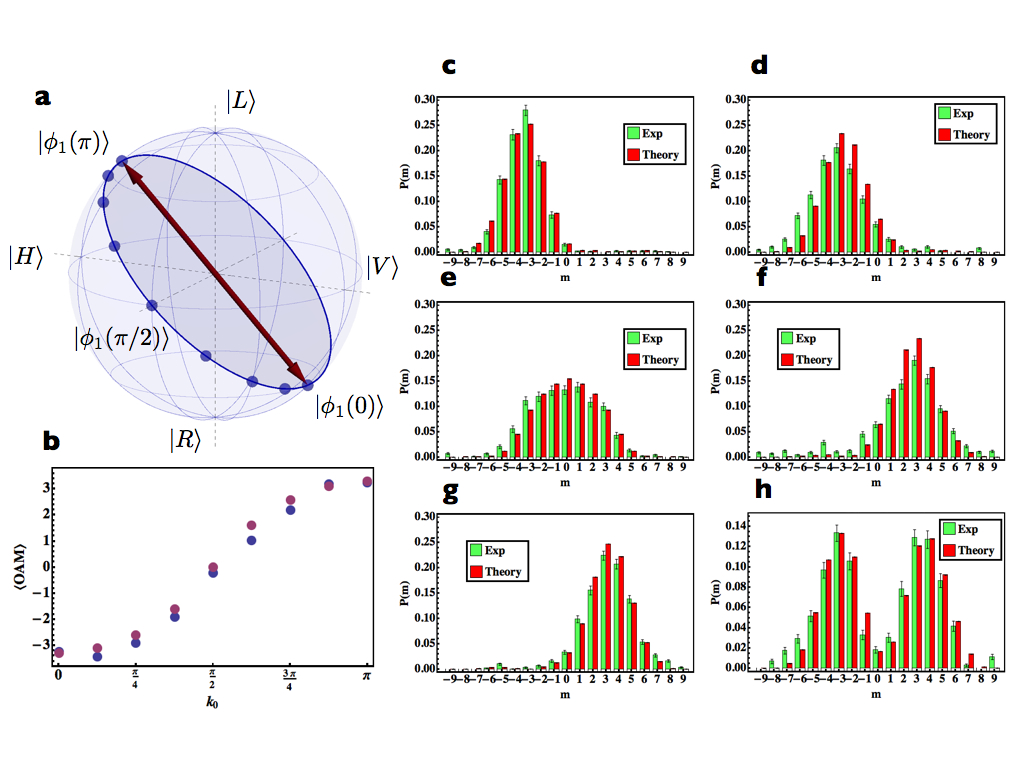}
\caption{Quantum walk wavepacket dispersion properties in the Brillouin zone. a) Poincar\'e sphere representation of the polarization (or SAM) eigenstates $\ket{\phi_1(k)}$ prepared in our experiments, for different values of the quasi-momentum $k$ in the irreducible Brilluoin zone $(0,\pi)$ taken in steps of $\pi/8$ (blue dots). These states lie on a maximal circle (blue line) of the sphere. b) Mean OAM after a five-steps QW for a single photon prepared in a Gaussian wavepacket with $\sigma=2$ and $s=1$, with different values of average quasi-momentum $k_0$ in the range $(0,\pi)$. Blue and purple points are associated to experimental data and theoretical predictions, respectively; Poissonian statistical uncertainties are too small to be shown in the graph. c-g) Final OAM distribution associated to some of these cases (summed over different polarizations). Panels refer to $k_0=0,\pi/4,\pi/2,3\pi/4,\pi$, respectively. h) OAM distribution after five-step QW for a wavepacket whose coin is prepared in the superposition state $(\ket{\phi_1(0)}+\ket{\phi_2(0)})/\sqrt 2$. As predicted by the theory, it splits into two components propagating in opposite directions, thus generating a maximally-entangled SAM-OAM state. In panels c-h) poissonian statistical uncertainties at plus-or-minus one standard deviation are shown by error bars. The similarity between experimental and theoretical OAM distributions are  $(98.9\pm0.2)\%$, $(96.2\pm0.4)\%$, $(98.4\pm0.3)\%$, $(93.2\pm0.6)\%$, $(99.1\pm0.2)\%$ and $(97.3\pm0.4)\%$, respectively.} 
\label{fig:dispersion}
\end{figure*}

In Fig.\ \ref{fig:packets}, we report the experimental ``real-time'' (i.e., step-by-step) observation of these propagating packets for a 5-step QW. These data refer in particular to the band $s=1$, with $k_0=\pi$ and $k_0=\pi/2$, corresponding to maximum and vanishing group velocities, respectively, with a step operator implemented by a QP plus a QWP. Next, we proceeded to explore the whole irreducible Brillouin zone by varying the average quasi-momentum $k_0$ in steps of $\pi/8$ across the $(0,\pi)$ range. At each value of $k_0$, in order to obtain a single wavepacket propagation, the SAM input state must be prepared in the eigenstate $\ket{\phi_1(k_0)}$, corresponding to a specific elliptical polarization. As a result of the so-called sublattice or chiral symmetry\cite{Kita10_PRA}, the corresponging SAM (or coin) eigenstates of these wavepackets describe a maximum circle in the Poincar\'e polarization sphere, as illustrated in Fig.\ \ref{fig:dispersion}a. The number of full rotations of the vector $\ket{\phi_1(k)}$ on the sphere, as $k$ varies from $-\pi$ to $\pi$, is a topological property of the QW system. In our case, we observe a single full rotation (we actually see half a rotation, as we tested only half Brillouin zone), thus verifying the topological class of our system. Other topological QW phases could be realized by modifying the QW step operator $\hat U$, as discussed in Ref.\ \onlinecite{Kita10_PRA}. We then determined the group velocity of these wavepackets by measuring the mean OAM exit value after 5 steps, as shown in Fig.\ \ref{fig:dispersion}b. The whole OAM distribution for some of these points is also shown in Figs.\ \ref{fig:dispersion}c-g.

Finally, the behavior of a wavepacket whose coin is prepared in the superposition state $(\ket{\phi_1(k_0)}+\ket{\phi_2(k_0)})/\sqrt 2$ was also investigated. As a result of the spin-orbit coupling, the wavepacket splits into two components propagating in opposite directions, as shown in Fig.\ \ref{fig:dispersion}h. In this example, the QW clearly leads to the generation of entanglement between the SAM and OAM degrees of freedom. The large average OAM separation obtained between the two wavepacket components implies that the obtained final photon state can be interpreted as a Schr\"odinger ``cat state'' in OAM space.

\subsection*{Two-photon quantum walk}
The experiments discussed above were carried out in the heralded single-photon regime. Although the latter is a quantum regime, it behaves equivalently to a classical one, as the resulting probability distributions are identical to the intensity distributions that would be obtained using classical (coherent) light. However, the OAM QW platform introduced in this work is also immediately suitable for simulating multiparticle quantum processes, for which quantum interferences cannot be reproduced classically.
%
%
\begin{figure*}[t]
\centering
\includegraphics[width=14cm]{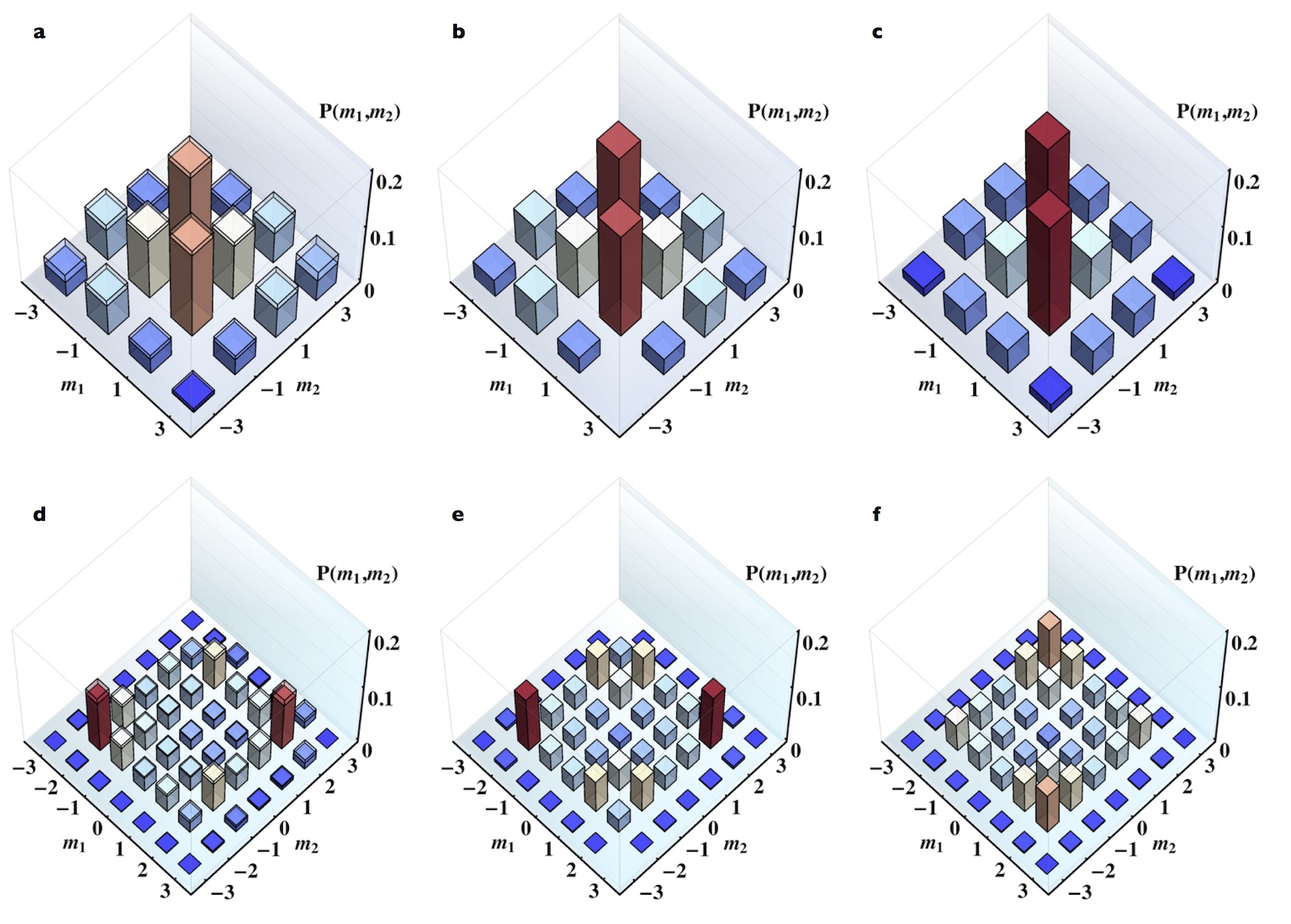}
\caption{Three-step quantum walk for two identical photons. In this case, only final OAM probabilities are shown (summed over different polarizations). a-c) Case of standard walk. a) Experimental results. Vertical bars represent estimated joint probabilities for the OAM of the two photons. Since the two measured photons detected after the BS splitting are physically equivalent, their counts are averaged together, so that $(m_1, m_2)$ and $(m_2, m_1)$ pairs actually refer to the same piece of data. Even values of $m_1$ and $m_2$ are not included, since they correspond to sites that cannot be occupied after an odd number of steps. b) Theoretical predictions for the case of indistinguishable photons. c) Theoretical predictions for the case of distinguishable photons, shown to highlight the effect of two-photon interference (Hong-Ou-Mandel effect) in the final probabilities. It can be seen that the experimental results agree better with the theory for indistinguishable photons. d-f) Case of hybrid walk (with $\delta=1.46$). d), e) and f) refer respectively to experimental data, indistinguishable photon theory and distinguishable photon theory, as in the previous case. The QW step in these two-photon experiments is implemented with a QP and a QWP. Again, our experiment is in good agreement with the theory based on indistinguishable photons, proving that two-photon interferences are successfully implemented in our experiment. The similarities between experimental and predicted quantum distributions (IPT model) are $(98.2\pm0.4)\%$ and $(95.8\pm0.3)\%$ for the standard and the hybrid walk, respectively. The similarities with the DPT model are instead $96.4\%$ and $91.8\%$, respectively. The color scale (common to all panels referring to the same case) reflects the vertical scale, to help comparing the patterns.}
\label{fig:data2}
\end{figure*}

To provide a first demonstration of this additional feature, we investigated the simultaneous QW of two identical photons. In this case, both photons generated in the SPDC process were sent in the QW setup, after adjusting their input polarization state to $\ket{R}\ket{L}$, selected as a typical case. At the exit of the QW cascade, we split the two photons with a beam splitter and analyze them both in polarization and OAM, so as to obtain their joint probability distribution (see Fig.\ \ref{fig:setup2p} for the experimental layout). In Fig.\ \ref{fig:data2}, the results relative to a 3-step QW with localized OAM input $m=0$ are reported and compared with the theoretical predictions obtained for indistinguishable photons (while taking into account the effect of the final beam splitter), hereafter labeled as ``indistinguishable-photon theory'' (IPT). The two distributions show a good quantitative agreement, as is confirmed by their similarities being higher than 95\% (see figure captions for details). These similarities are defined as in the single photon case, with the index $m$ replaced by the pair of OAM values $(m_1,m_2)$. The predicted distributions for the case of distinguishable photons (DPT) are also shown for comparison, to highlight the role of two-particle interference in the final distributions. The similiarities of the data with the DPT distributions are significantly lower. However, the similarity is not a very sensitive test, as it tends to remain high even for fairly different distributions. Hence, we also computed the ``total variation distance'' (TVD, defined as the sum of the absolute values of all probability differences divided by two) for the two cases. In the standard case, the TVD of the experimental distribution with the IPT one is $(6.5\pm0.9)\%$, to be compared with the TVD of $(16.5\pm0.9)\%$ for the DPT model. In the hybrid case, the TVD with the IPT is $(13.5\pm0.7)\%$, to be compared with $(21.1\pm0.7)\%$ for the DPT. These values confirm that two-photon interferences are present in our experiment. We ascribe the residual discrepancies between the observed distributions and the IPT quantum predictions to systematic errors arising from imperfect alignment of the setup.

On the other hand, it is also possible to demonstrate a quantum behavior in the observed distributions independently of any specific model for the photon propagation in the QW system, so as to be insensitive to alignment imperfections or other kinds of systematic errors. As discussed in the SM, this is accomplished by testing the violation of certain characteristic inequalities that constrain any possible correlation distribution obtained with two classical light sources instead of two photons \cite{Peru10_Sci}, or with two distinguishable photons. The measured distributions indeed violate these inequialities by several standard deviations, as illustrated in the SM (Figs.\ S6 and S7). This proves once more that the measured correlations must be quantum and that they include the effect of multiparticle interference.

\section*{Discussion}
In this article, we have demonstrated a single- and multi-photon quantum walk simulator based on single beam propagation through linear optical devices. The realized architecture is efficient and stable. Moreover, in contrast to other photonic QW implementations, the number of optical components employed scales only linearly with the number of steps, since at each step all OAM values are addressed simultaneously by a single optical element, whose transverse extension remains constant at each step. It must be noted however that this advantage in scaling remains valid only as long as the entire QW takes place in the optical near-field, where the beam cross-section size will remain approximately constant, while in the far-field the transverse size of the optical components will have to increase with the OAM range (see the SM).

An important advantage of this platform is the possibility to prepare the walker initial state, even if extended over many lattice sites, with high accuracy and flexibility. We exploited this feature to investigate the effective band structure of QWs, demonstrating the propagation of Gaussian wavepackets for different points in the Brillouin zone and exploring the associated topological structure arising from the spin-orbit coupling. For a certain initial state, the wavepacket is split in two by the QW evolution, leading to a quantum ``Schr\"odinger cat state'' in OAM. In prospect, it will be very interesting to simulate the quantum propagation of extended states of two (or more) photons, possibly entangled to each other, such as those naturally generated in the SPDC process. Moreover, engineering the initial state of the walker is a possible strategy for the simulation of complex quantum walk dynamics through the combination of suitable delocalized initial conditions plus a standard QW evolution; a theoretical proposal was reported recently for the case of ``driven quantum walks'' \cite{Hami14_PRL}.

A current limitation of our approach is that the walk evolution cannot be position-dependent (that is, OAM-dependent), in contrast to other implementations\cite{Schr12_Sci,Cres13_NatPhot}. This limitation could be overcome in the future by introducing additional optical elements acting on the azimuthal coordinate (for example, a Dove's prism can introduce an OAM-dependent phase shift) or by exploiting the radial beam coordinate, which couples with OAM in free propagation and can be acted on by a radially-patterned optical element. On the other hand, our approach allows a very convenient and easy control of the evolution operator at each step, including the possibility of fully-automated fast switching of its properties by introducing electrooptical devices to manipulate the polarization or by electrically controlling the q-plate tuning. This may enable, for example, the simulation of a quantum system having a time-dependent Hamiltonian or that of a statistical ensemble of quantum systems with different Hamiltonians. Another potential advantage of the present implementation is the possibility to carry out a full quantum tomography of the outgoing state, which is very challenging for standard interferometric implementations. Finally, we must mention the important limitation of our platform, common to all fully photonic QW implementations, of not being able to simulate particle interactions. A possible future strategy to overcome this limitation might be based on ideas similar to those proposed by Knill, Laflamme and Milburn for doing quantum computation with linear optics \cite{Knill01_Nat,Vite13_NatPhot}.

\newpage



\vspace{1 EM}
\noindent\textbf{Acknowledgments}\\
\noindent We thank Pei Zhang for an early suggestion of the possibility to carry out a photonic quantum walk in OAM following the scheme proposed in his paper and Antonio Ramaglia and Marco Cilmo for lending some equipment. This work was partly supported by the Future Emerging Technologies FET-Open Program, within the 7$^{th}$ Framework Programme of the European Commission, under Grant No.\ 255914, PHORBITECH. F.S. acknowledges also ERC Starting Grant 3D-QUEST (grant agreement no. 307783). E.K. and R.W.B. acknowledge the support of the Canada Excellence Research Chairs (CERC) Program.
\vspace{1 EM}

\noindent\textbf{Author Contributions}\\ 
\noindent F.C., F.M, E.K., F.S., E.S., R.W.B and L.M. devised various aspects of the project and designed the experimental methodology. F.C., F.M. and H.Q., with contributions from E.K., D.P., C.d.L., carried out the experiment and analyzed the data. S.S. prepared the q-plates. F.C., F.M. and L.M. wrote the manuscript, with contributions from E.K. All authors discussed the results and contributed to refining the manuscript.
\vspace{1 EM}

\noindent\textbf{Competing interests}\\
 The authors declare no competing financial interests. Correspondence and requests for materials should be addressed to L.M. (lorenzo.marrucci@unina.it) or R.W.B. (boydrw@mac.com).
\newpage
\onecolumngrid
\appendix

\textbf{Supplementary Materials for Quantum walks and wavepacket dynamics on a lattice with twisted photons}
\vspace{1 EM}
\section{Effective band structure of the quantum walk system}
Due to the translational symmetry of the quantum walk (QW) system, eigenstates of the single step evolution operator $\hat U$ are obtained as the direct product of quasi-momentum eigenstates $\ket{k}_w$ in the walker space $\mathcal H_w$ with suitable eigenvectors $\ket{\phi_s(k)}_c$ living in the coin (polarization) space $\mathcal H_c$ (in the following, subscripts $w$ and $c$ will be omitted whenever there is no risk of ambiguity). We label these eigenstates as $\ket{k,s}=\ket{\phi_s(k)}_c\otimes\ket{k}_w$, with $s\in\{1,2\}$\cite{Kita10_PRA,Kita12_NatCom}. The corresponding eigenvalues are $\lambda_s(k)=e^{-i \omega_s(k)}$, where the relation between the quasi energy $\omega$ and the quasi momentum $k$ for a  coin operator $\hat {T}$ having $a=b=1/\sqrt{2}$ (the standard Hadamard gate) has the following analytical expression\cite{Abal06_PRA}:
\begin{equation}\label{met:omega}
\omega_2(k)=\arcsin{\left(\frac{\sin k}{\sqrt 2}\right)};\quad \omega_1(k)=\pi-\omega_2(k).
\end{equation}
In the SAM-OAM implementation of the QW process, this specific dispersion relation is obtained when using a quarter-wave plate and a q-plate in each step. Both energy and momentum are periodic quantities, as a result of the discrete nature of the space-time considered in the process. In Fig.\ \ref{fig:dispersion} we report a graph for the dispersion relation and the group velocity dispersion. The two bands $s=1$ and $s=2$ are characterized by a finite energy gap, and their group velocities, defined as $V_s=d\omega_s/dk$, have the same magnitude but opposite sign, i.e.\ $V_1(k)=-V_2(k)$. As shown in the graph, $V_s$ is bounded in the range $(-1/\sqrt 2,1/\sqrt 2)$, and it vanishes for $k=\pm\pi/2$. As shown in Fig.\ 5 of the main article, the coin eigenstates $\ket{\phi_s(k)}_c$ span a great circle on the Poincar\'e sphere, as the result of the so-called chiral or sublattice symmetry\cite{Kita10_PRA}.
\begin{figure*}[h]
\centering
\includegraphics[width=14cm]{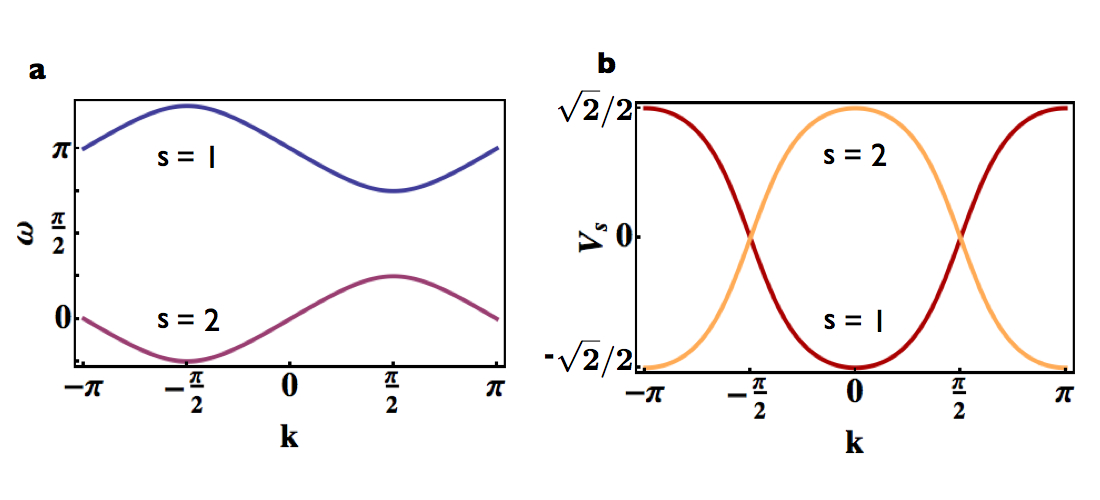}
\caption{Band structure of the QW system. (a) The plot shows the dispersion relation $\omega_s(k)$ for both bands $s=1$ and $s=2$. A finite energy gap can be observed. (b) Dispersion of the group velocity $V_s(k)$. It's worth noticing that when $k=\pm\pi/2$, the group velocity vanishes for both bands and, at the same time, it has the maximum slope. In the context of wavepackets dynamics, for these values of the quasi-momentum the dispersion of the group velocity will give a larger contribution to the broadening of the initial envelope in comparison with the case of packets propagating at non-zero speed. 
}
\label{fig:dispersion}
\end{figure*}
\section{The q-plate device}
A q-plate (QP) consists of a thin slab of uniaxial birefringent nematic liquid crystal sandwiched between two glasses, whose optical axis in the slab plane is engineered in an inhomogeneous pattern, according to the relation
\begin{equation}\label{met:oa}
\alpha(\phi)=q\,\phi+\alpha_0,
\end{equation}
where $\alpha$ is the angle formed by the optical axis with the reference (horizontal) axis, $\phi$ is the azimuthal coordinate in the transverse plane of the device, $q$ is the topological charge of the plate and $\alpha_0$ is the axis direction at $\phi=0$. When light passes through a QP, the angle $\alpha_0$ controls the relative phases of the different OAM components in the output state. For arbitrary $\alpha_0$, the action of the QP is described by the following equations
\begin{align}
\widehat Q^{\alpha_0}_\delta\ket{L,m}&=\cos{(\delta/2)}\ket{L,m}-i\sin{(\delta/2)}e^{i\,2\alpha_0}\ket{R,m+2q},\cr 
\widehat Q^{\alpha_0}_\delta\ket{R,m}&=\cos{(\delta/2)}\ket{L,m}-i\sin{(\delta/2)}e^{-i\,2\alpha_0}\ket{L,m-2q},
\end{align}
which reduces to Eq.\ 4 of the main text when $\alpha_0=0$. A vanishing relative phase between the two OAM-shifted terms is required to properly implement the operator $\hat U$ describing the QW process. To achieve this, all QPs in our setup were oriented so as to match the condition $\alpha_0=0$.
\section{Role of the radial modes and Gouy phases}
Our QW realization relies on the encoding of the walker state in the transverse modes of light, in particular exploiting the azimuthal degree of freedom. For simplicity, the radial structure of the mode is not considered explicitly in our scheme. However, a full treatment of the optical process requires one to take the radial effects into account. Indeed, all optical devices used to manipulate the azimuthal structure and hence the OAM of light, including the QP, unavoidably introduce some alteration of the radial profile of the beam, particularly when subsequent free propagation is taken into account.

In this context, we choose Laguerre-Gauss (LG) modes as the basis, since they provide a set of orthonormal solutions to the paraxial wave equation. LG modes are indexed by an integer $m$ and a positive integer $p$ which determine the beam azimuthal and radial structures, respectively. Using cylindrical coordinates $r,\phi,z$, these modes are given by 
\begin{align}
	\label{eq:lg}
	\mbox{LG}_{p,m}(r,\phi,z)&=\sqrt{\frac{ 2^{|m|+1}p!}{\pi w(z)^2\,(p+|m|)!}}\,\left(\frac{r}{w(z)}\right)^{|m|} e^{-\frac{r^2}{w(z)^2}} L_{p}^{|\ell|}\left(\frac{2r^2}{w(z)^2}\right)\,e^{\left(\frac{i\pi r^2}{\lambda R(z)}\right)}\,e^{im\phi}\,e^{-i(2p+|m|+1)\arctan{\left(\frac{z}{z_R}\right)}},
\end{align}
where $\lambda$ is the wavelength, $w(z)=w_0\,\sqrt{1+(z/z_R)^2}$, $R(z)=z\left[1+(z/z_R)^2\right]$ and $z_R=\pi w_0^2/\lambda$ are the beam radius, wavefront curvature radius and Rayleigh range, respectively, $w_0$ being the radius at the beam waist~\cite{Siegman}. $L_{p}^{|\ell|}(x)$ are the generalized Laguerre polynomials.

As already discussed, the QP raises or lowers the OAM content of the incoming beam, according to its polarization state. Due to presence of the singularity at the origin, the QP also alters the radial index of the incoming beam. The details of these calculations are reported in Ref.\ \onlinecite{Kari09_OL}. Based on this analysis and assuming a low birefringence of the liquid crystals, a \emph{tuned} QP (i.e.\ with $\delta=\pi$) transforms a circularly polarized, e.g.\ left-handed, input LG$_{0,m} (r,\phi,0)$ beam as follows:
\begin{align}\label{eq:qplate}
\widehat Q_{\pi}\mbox{LG}_{0,m}(r,z)\ket{L,m}&=-i\,\text{HyGG}_{|m|-|m+1|,m+1}(r,z)\ket{R,m+1},
\end{align}
where LG$_{0,m} (r,0)$ (without the $\phi$ variable) denotes the radial part of the LG$_{0,m} (r,\phi,0)$ mode (i.e.\ with $\phi=0$), HyGG$_{p,m} (r,z)$ stands for the amplitude of Hypergeometric-Gauss (HyGG) modes~\cite{Kari07_OL} and the azimuthal term $e^{im\phi}$ has been replaced by the ket $\ket{m}$.  Introducing dimensionless coordinates $\rho=r/w_0$ and $\zeta=z/z_R$, these modes are given by
\begin{eqnarray}\label{eq:HyGG}
   \hbox{HyGG}_{pm}(\rho,\zeta)&=&i^{|m|+1}\sqrt{\frac{2^{p+|m|+1}}{\pi\Gamma(p+|m|+1)}} \frac{\Gamma\left(1+|m|+\frac{p}{2}\right)}{\Gamma\left(|m|+1\right)}\,\\
    &{\ensuremath\times}&\zeta^{\frac{p}{2}}(\zeta+i)^{-(1+|m|+\frac{p}{2})} \rho^{|m|}\,e^{-\frac{i\rho^2}{(\zeta+i)}}
   {}_1\!F_1\left(-\frac{p}{2},1+|m|;\frac{\rho^2}{\zeta(\zeta+i)}\right)\nonumber
\end{eqnarray}
where $\Gamma(x)$ is the gamma function and $F_1(a,b;x)$ is a confluent hypergeometric function. In order to determine the radial mode alteration introduced by the QP, we can expand the output beam in the LG modes basis, i.e.\ $\text{HyGG}_{|m|-|m+1|,m+1}=\sum_p{c_p\text{LG}_{p,m+1}}$~\cite{Kari07_OL}. The expansion coefficients are given by
\begin{eqnarray}\label{eq:coeff}
c_p=\sqrt{\frac{1}{p!\,m!\,\left(p+|m+1|\right)!}}\,\frac{\left(|m+1|+|m|\right)!\,\Gamma\left(p+\frac{|m+1|-|m|}{2}\right)}{\Gamma\left(\frac{|m+1|-|m|}{2}\right)}
\end{eqnarray}
\begin{table}[tb]
  \centering
  \caption{\label{tab:tab1} Power coefficients of the various $p$-index terms appearing in the expansion of the beam emerging from a QP (with $q=1/2$) in the LG-mode basis, assuming that the input is an $L$-polarized LG mode with $p=0$ and the given OAM $m$ value.}
 \begin{tabular*}{0.4\textwidth}{@{\extracolsep{\fill} }  c|cccc}
    \hline\hline
    OAM          &     $|c_0|^2$   & $|c_1|^2$  &     $|c_2|^2$   & $|c_3|^2$ \\ \hline\hline
    $m=0$       &   0.785  &  0.098  &  0.036   &  0.019 \\ \hline
    $m=1$       &   0.883  &  0.073  &  0.020   &  0.008 \\ \hline
    $m=2$       &   0.920  &  0.057  &  0.012   &  0.004 \\ \hline
    $m=3$       &   0.939  &  0.046  &  0.008   &  0.002  \\
  \hline\hline
\end{tabular*}
\end{table}
Table~\ref{tab:tab1} shows the squared coefficients of this expansion for input beams possessing different OAM values. As can be seen, the effect of the QP on the radial mode decreases for beams having higher OAM values, so that only the $p=0$ coefficient that was already present at the input retains a large value after the QP, and one can approximately neglect higher-$p$ terms. If the final detection based on coupling in a single-mode fiber filters only this term, then the presence of the other terms only introduces a certain amount of losses in the system. Hence, within such approximation, the $p$ quantum number plays essentially no role and it can be ignored (except for the Gouy phase, which is discussed further below).

Even stronger is the argument one can use if the entire QW simulation takes place in the optical near field. Indeed, at the pupil plane $(\zeta\rightarrow0)$ the expression for the amplitude of HyGG and LG modes simplifies to
\begin{eqnarray}\label{eq:HyGGpupil}
	\hbox{LG}_{p'm'}(\rho,0)&\propto&L_{p}^{|\ell|}(\rho^2)\rho^{|m|}e^{-\rho^2}\\
	\hbox{HyGG}_{pm}(\rho,0)&\propto&\rho^{p+|m|}e^{-\rho^2}.\nonumber
\end{eqnarray}
Combining Eq.\ \ref{eq:qplate} and Eq.\ \ref{eq:HyGGpupil}, it is straightforward to prove that the action of a QP placed at the pupil plane of the beam is given by
\begin{align}\label{eq:qplatepupil}
\widehat Q_{\pi}\mbox{LG}_{0,m}(\rho,0)\ket{L,m}&=-i\,\mbox{LG}_{0,m}(\rho,0)\ket{R,m+1}.
\end{align}
In other words, at the immediate output of the device, the QP ideally results only in the increment of the OAM content, without any alteration of the radial profile. This result remains approximately valid as long as the beam is in the near field, that is for $\zeta\ll1$, except for a region very close to the central singularity and for some associated fringing that occurs outside the singularity. Both these effects can be neglected for $\zeta\ll1$, as the overlap integral of the resulting radial profile with the input Gaussian profile remains close to unity (for example, at $\zeta=0.1$ this overlap is still about 0.93 for a HyGG mode with $m=1$). We exploit this property to minimize any effect due to a possible coupling between the azimuthal and the radial degree of freedom introduced by the QP. The setup was built in order to have all the steps of the QW in the near field of the input photons. To achieve this, we prepared the beam of input photons to have $z_R>10$ m, while the distance between the QW steps was $d\approx10^{-2}z_R$. For realizing a QW with high number of steps, a lens system could be used to image the output of each QW unit at the input of the next one; in this way the whole process may virtually occur at the pupil, i.e.\ at $\zeta=0$, thus effectively canceling all radial-mode effects.

Free space propagation of photonic states carrying OAM is characterized by the presence of a phase term, usually referred to as Gouy phase, that evolves along the optical axis. Considering for example LG states of Eq.~\ref{eq:lg}, this phase factor is given by $\exp{\left[-i(2p+|m|+1)\arctan{(z/z_R)}\right]}$, where $z$ is the coordinate on the optical axis with respect to the position of the beam waist. The different phase evolution occurring for different values of $|m|$ could be a significant source of errors in the QW implementation. Let us assume that after step $n$ in the QW setup the state of the photon is $\ket{\psi}=\sum_m{c_m\ket{m}}$, where for simplicity we consider only modes with $p=0$. When entering the following step, the coefficients $c_m$ will evolve to $c_m'=e^{-i2|m|\arctan{(d/z_R)}}c_m$, where $d$ is the distance between two steps along the propagation axis. At the step $n+1$, coefficients $c_m$ and $c_m'$ lead to different interferences between the OAM paths, altering the features of the QW process. In our implementation we made this effect negligible by relying on the condition $d/z_R\ll1$: indeed, as discussed previously, in our setup we had that $z_R>10$ m and $d\simeq10$ cm. Again, an alternative strategy could be based on using a lens system to image each QP on the following one; at image planes all relative Gouy phases vanish.

Let us conclude this Section by noting that the need to remain in the optical near field is not a ultimate scaling limitation for our QW process implementation, as the Rayleigh range $z_R$ can be made as large as desired by increasing the beam waist $w_0$. The beam waist $w_0$ scales as the square root of $z_R$. Hence, the overall needed resources of our QW implementation, defined as the number of needed optical devices multiplied by their transverse area, scales linearly in the number of steps $n$ as long as $w_0$ can remain constant, while there is a crossover to the standard quadratic scaling when $z_0$ must be further increased.

\section{Preparation and measurement of the OAM states}
Our approach based on encoding the walker system in the OAM degree of freedom of a single photon enables us to easily prepare both localized and ``delocalized'' initial states, the latter corresponding to superpositions of multiple lattice sites.
\begin{figure*}[htb]
\centering
\includegraphics[width=16cm]{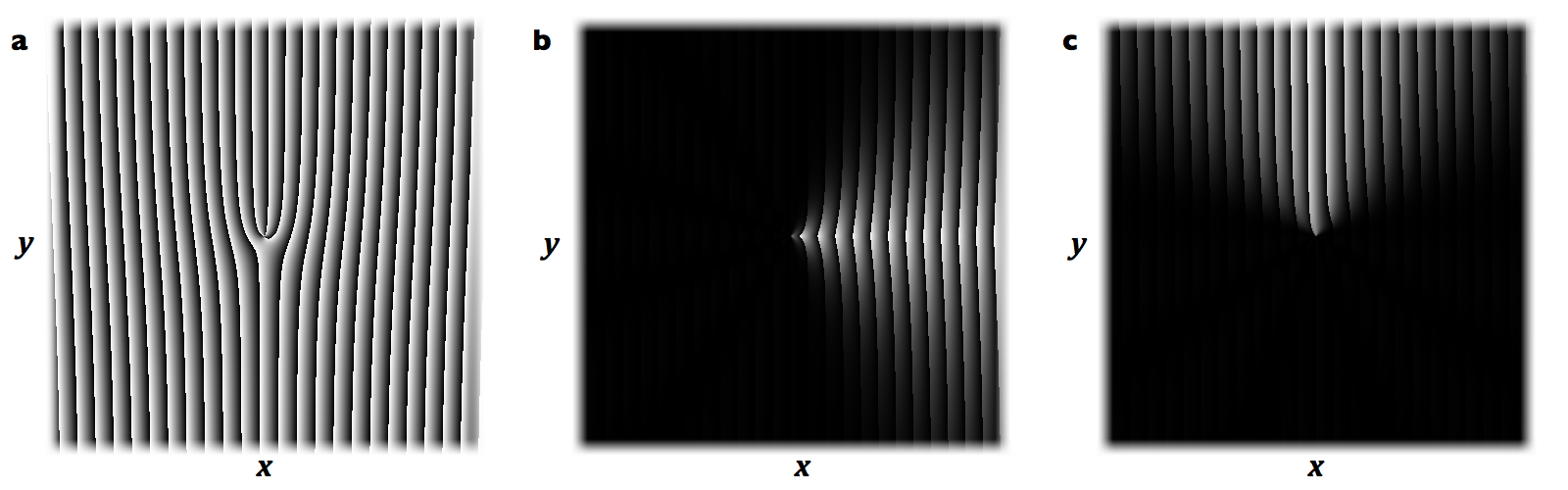}
\caption{Holograms for the preparation of the OAM initial state before the QW process. In the pictures the phase $\mathcal F(x,y)$, ranging in the interval $(0,2\pi)$, is encoded in the grayscale level of each pixel. Panels (a)-(c) refer to different initial states in the OAM space. (a) Localized initial state $\ket{\psi_0}_w=\ket{+3}_w$. (b)-(c) Delocalized Gaussian wavepackets  $\ket{\psi_0}_w=A_0\sum_m {e^{-ik_0m}e^{-m^2/2\sigma^2}\ket{m}_w}$~, with $\sigma=2$ and $k_0=0$ and $k_0=\pi/2$, respectively. $A_0$ is a normalization constant.} 
\label{fig:holo}
\end{figure*}
This is accomplished by means of a holographic technique\cite{Bold13_OL,Damb13_SciRep}, which provides an exact solution to the problem of generating arbitrary transverse distributions for a paraxial optical field. The relation which links the transverse component of the generated field, described by $E(x,y)=\mathcal A(x,y) e^{i\mathcal P(x,y)}$, and the phase $\mathcal F(x,y)$ to be introduced by the hologram (for a plane-wave input) is given by
\begin{align}\label{eq:holo}
	\mathcal F=\mathcal M(\mathcal A)\hbox{Mod}{[(\mathcal P+\mathcal B-\pi \mathcal M(\mathcal A)),2\pi]},
\end{align}
where $\mathcal B(x,y)$ corresponds to a blazed grating, which defines the diffraction direction, and $\mathcal M(\mathcal A)=(1+1/\pi\,\text{Sinc}^{-1}(\mathcal A))$, where $\text{Sinc}^{-1}$ takes values in $[-\pi,0]$.

The phase hologram is displayed on a spatial light modulator (SLM); when a plane wave is impinging on this device, the field $E(x,y)$ is generated at the first diffraction order, with an efficiency depending strongly on the phase and the amplitude distributions $(\mathcal A,\mathcal P)$. In our case, we used a Gaussian beam as input, which leads to the same azimuthal field distribution as for a plane-wave. As an example, in Fig.\ref{fig:holo} we report the holograms computed for the generation of both localized and delocalized (i.e.\ Gaussian wavepacket) initial states. It can be noticed that the average quasi-momentum $k_0$ corresponds also to the average azimuthal angle in real space for the optical mode.

In order to measure the OAM value of the photons, we have implemented the widely used holographic technique introduced by Mair \emph{et al.} in 2001~\cite{Mair01_Nat}. In this technique, the helical phase-front of the optical beam is ``flattened'' by diffraction on a pitch-fork hologram (displayed on another SLM) and the Gaussian component of the beam at the far-field is then selected by a single mode optical fiber (SMF). This approach, as shown in Ref.\ \onlinecite{Hamm14_JOSAB}, leads to a biased outcome for the different OAM values, since the coupling efficiency of this projective measurement changes according to the OAM of the input beam. This issue is analyzed in detail in Ref.\ \onlinecite{Hamm14_JOSAB}. We have taken this effect into account by measuring experimentally the coupling efficiency for different OAM values and then correcting the corresponding measured probabilities.

In the case of two photons, the OAM measurement was carried out in the same way, by splitting the beam with a non-polarizing symmetrical beam splitter (BS), projecting the two output beams onto two distinct holograms displayed simultaneously on two portions of the SLM, and then coupling both diffracted beams into single-mode fibers.

In prospect, a more efficient OAM measurement approach for these kind of QW simulations (although perhaps less convenient for full quantum-state tomography) could be based on using the OAM-sorter devices of the kind introduced in Refs.\ \onlinecite{Berk10_PRL,Mirh13_NatCom}.

\section{Experiment: supplementary information}
In the case of a single photons, we have carried out measurements with a few other choices of input polarization, besides those already shown in the main article. The results are reported in Fig.\ \ref{fig:data1}.
\begin{figure*}[htb]
\centering
\includegraphics[width=18cm]{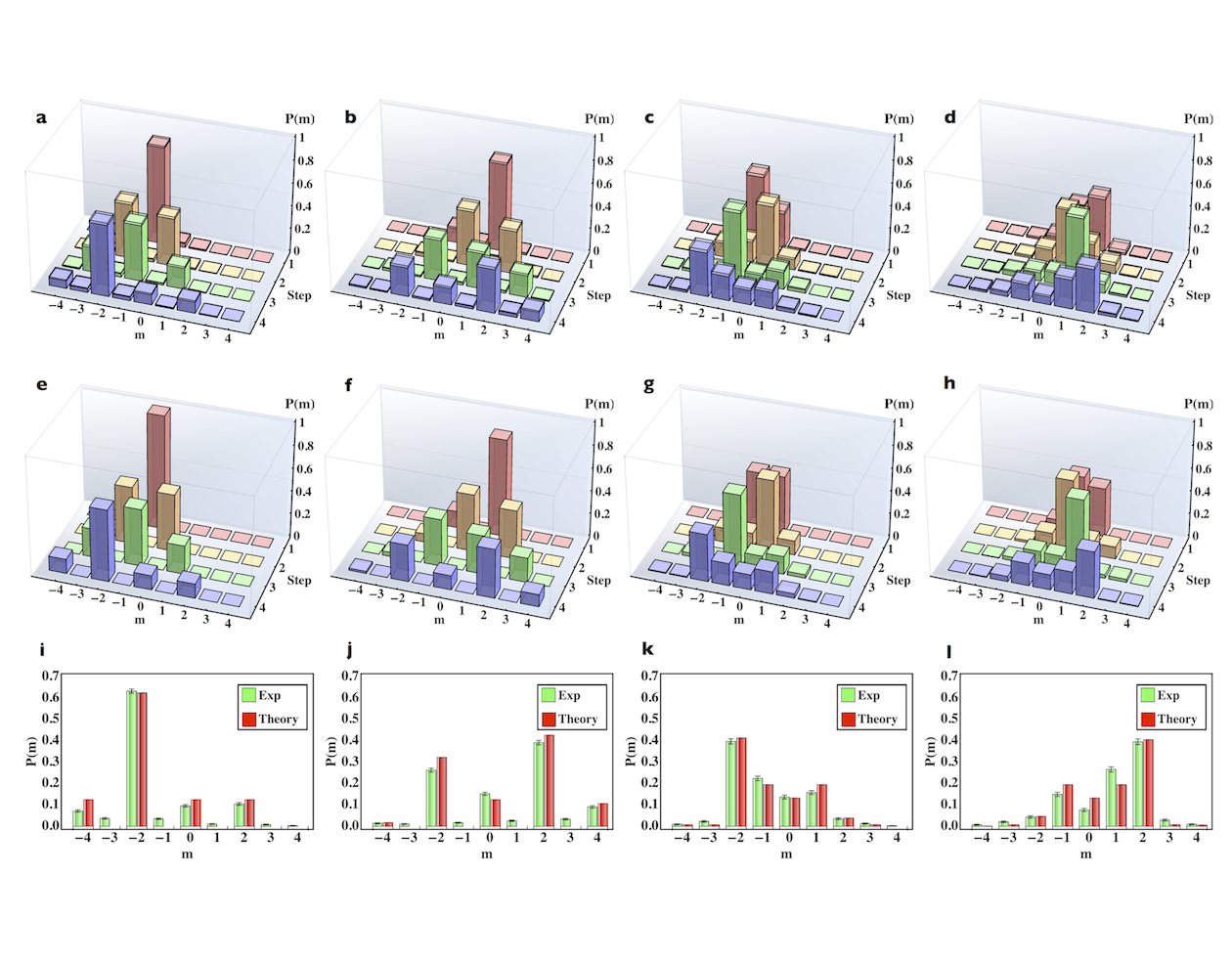}
\caption{Supplementary data for the four-step quantum walk for a single photon, with various input polarization states. (a)-(d) Experimental results, including both intermediate and final probabilities for different OAM states in the evolution (summed over polarizations). The intermediate probabilities at step $n$ are obtained by switching off all QPs that follow that step, that is setting $\delta=0$. Panels (a) and (b) refer to the standard case with two different input states for the coin subsystem, $(\alpha,\beta)= (1,-1)$ and $1/\sqrt{2} (1/\sqrt 2,1-i/\sqrt2)$, respectively. (c) and (d) refer to the hybrid case for $\delta=\pi/2$, with the coin subsystem, $(\alpha,\beta)= (1,-1)$ and $1/\sqrt{2} (1-i/\sqrt2,1/\sqrt 2)$, respectively. (e)-(h) Corresponding theoretical predictions. (i)-(l) Comparison of measured and predicted final probabilities. Poissonian statistical uncertainties at plus-or-minus one standard deviation are shown as error bars in panels (i)-(n) and as transparent-volumes in panels (a)-(e). The similarities between experimental and predicted OAM distributions are $(89.7\pm0.2)\%$, $(90.9\pm0.6)\%$, $(98.9\pm0.1)\%$ and $(97.0\pm0.4)\%$, respectively. Panels in the same column refer to the same configuration and initial states.}
\label{fig:dataSM}
\end{figure*}
\begin{figure*}[htb]
\centering
\includegraphics[width=8cm]{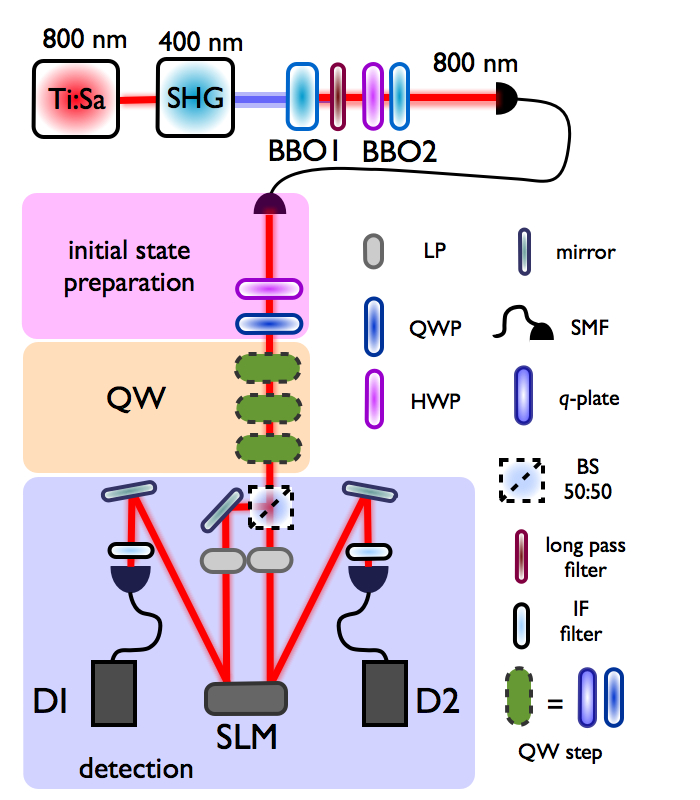}
\caption{Two-photon quantum walk apparatus. At the exit of the input SMF, a biphoton state is prepared by means of a QWP-HWP set in the state $\ket{L,R}$. Since in this case we explored only the case where the walk starts in $m_1=m_2=0$, the first SLM was not needed and was removed. The two photons, propagating along the same optical axes, go through a 3 steps QW. At the exit of the last step, a 50:50 BS randomly separates the two photons. At the exit of the BS, each arm is provided with a linear polarizer for the projective measurement in the SAM space. The OAM projection is then performed using an SLM and a SMF. For the projection on both arms, a single SLM was used, dividing its screen into two sections and showing independent holograms. Two interferential filters (IF) were used to filter the photon band so as to enhance the wavelength indistinguishability of the two photons. Finally, signals from photodiodes D1 and D2 provided the coincidence counts.}
\label{fig:setup2p}
\end{figure*}
\begin{figure*}[t]
\centering
\includegraphics[width=8cm]{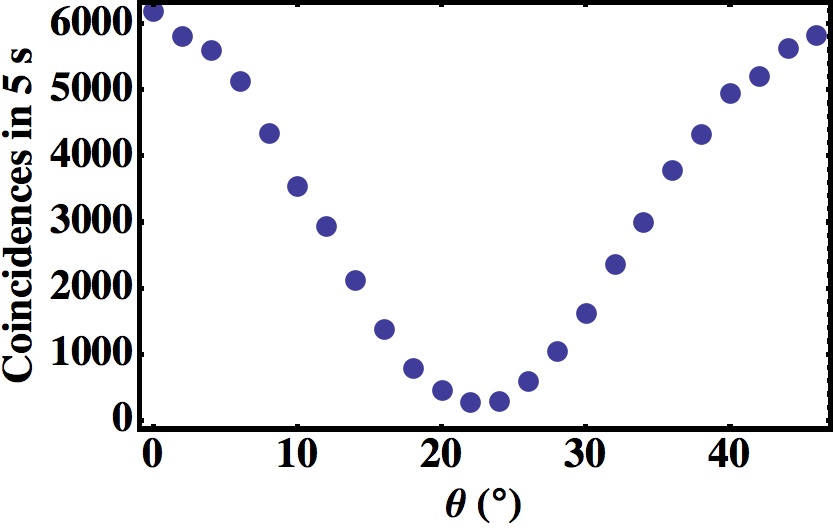}
\caption{Experimental verification of the indistinguishability of the two photon source through polarization Hong-Ou-Mandel (HOM) interference. The photons pairs in the polarization state $\ket{H,V}$ are sent through a HWP (oriented at an angle $\theta$ with respect to the horizontal direction) and a PBS. Coincidences counts are recorded at the exit of the two output ports of the PBS. When $\theta=22.5^{\circ}$, after the HWP the biphoton state becomes $1/\sqrt 2(\ket{H,H}-\ket{V,V})$. In this configuration, coincidence counts vanish as a result of the bunching of the two photons in the same polarization state. The measured visibility of the HOM fringes, for optimal temporal overlap (as obtained for proper walk-off compensation), is larger than 90\%. Poissonian statistical uncertainties at plus-or-minus one standard deviation are smaller than the size of circles representing the experimental points.}
\label{fig:HOM}
\end{figure*}

In Fig.\ \ref{fig:setup2p} we report the layout of the setup used for the simulation of a 2-particle QW. At the input, both photons exiting the SMF were sent through the QW step sequence, by removing the initial polarizing beam splitter (PBS). The biphoton state $\ket{H,V}$ generated by type-II spontaneous parametric down conversion (SPDC) is then converted into the state $\ket{L,R}$ by means of a quarter-wave-plate half-wave-plate (QWP-HWP) set. Since in this case we explored only the case where the walk starts in $m_1=m_2=0$, we removed the first SLM to improve the setup efficiency. The two photons, propagating along the same optical axes, go through a 3 steps QW. Since the two photons cannot be distinguished (except for the input polarization, which is however modified in the QW process) and propagate along the same axis, at the exit of the last step we introduced a 50:50 beam splitter (BS) to split them and perform independent SAM-OAM projective measurements on the two arms. This splitting stage and the duplication of the projection devices represent the main difference of this setup with respect to the apparatus discussed in Fig.\ 2 of the main article. This process has an efficiency of 50\%, as a result of the 1/2 probability that the two photons will exit from distinct BS ports. At the exit of the BS, each arm is provided with a linear polarizer for the projective measurement in the SAM space. As in the previous case, the OAM projection is performed using an SLM and a SMF. For the projection on both arms, a single SLM was used, dividing its screen into two sections and showing independent holograms. Before the last SMFs two interferential filters (IF) centered at 800 nm and with a bandwidth of 3.6 nm were used to filter the photon band so as to enhance the wavelength indistinguishability of the two photons. Finally, signals from photodiodes D1 and D2 were analyzed using a digital logic unit (time window 8 ns) combined with digital counters in order to get the final coincidence counts.

Before starting the main experiments, the indistinguishability of the two photons generated in the SPDC process was optimized and verified by carrying out a polarization Hong-Ou-Mandel two-photon interference, as shown in Fig.\ \ref{fig:HOM}. The results shown in the figure were obtained for optimal compensation of the walk-off occurring in the SPDC BBO crystal.

\section{Test of photon correlation inequalities}
Let us consider two photons entering the QW apparatus in fixed states 1 and 2. Here, we use a notation in which the state label at input/output includes both the OAM and the polarization. In our experiment, labels $1,2$ correspond to a vanishing OAM and $L,R$ polarizations. The output states $p$ will denote the combination of the OAM value $m$ and horizontal or vertical linear polarizations $H,V$. The unitary evolution of each photon from these input states to the final states can be described by a matrix $U_{l',l}$, where the first index corresponds to the input state and the second to the output one (notice that here we are making no assumptions on this matrix, except for unitarity). Hence, the QW evolution can be described by the following operator transformation law
\begin{equation}
\hat{a}^{\dag}_{l'} \rightarrow \hat{b}^{\dag}_{l'}=\sum_l U_{l',l}\hat{a}^{\dag}_l
\end{equation}

\begin{figure*}[t]
\centering
\includegraphics[width=18cm]{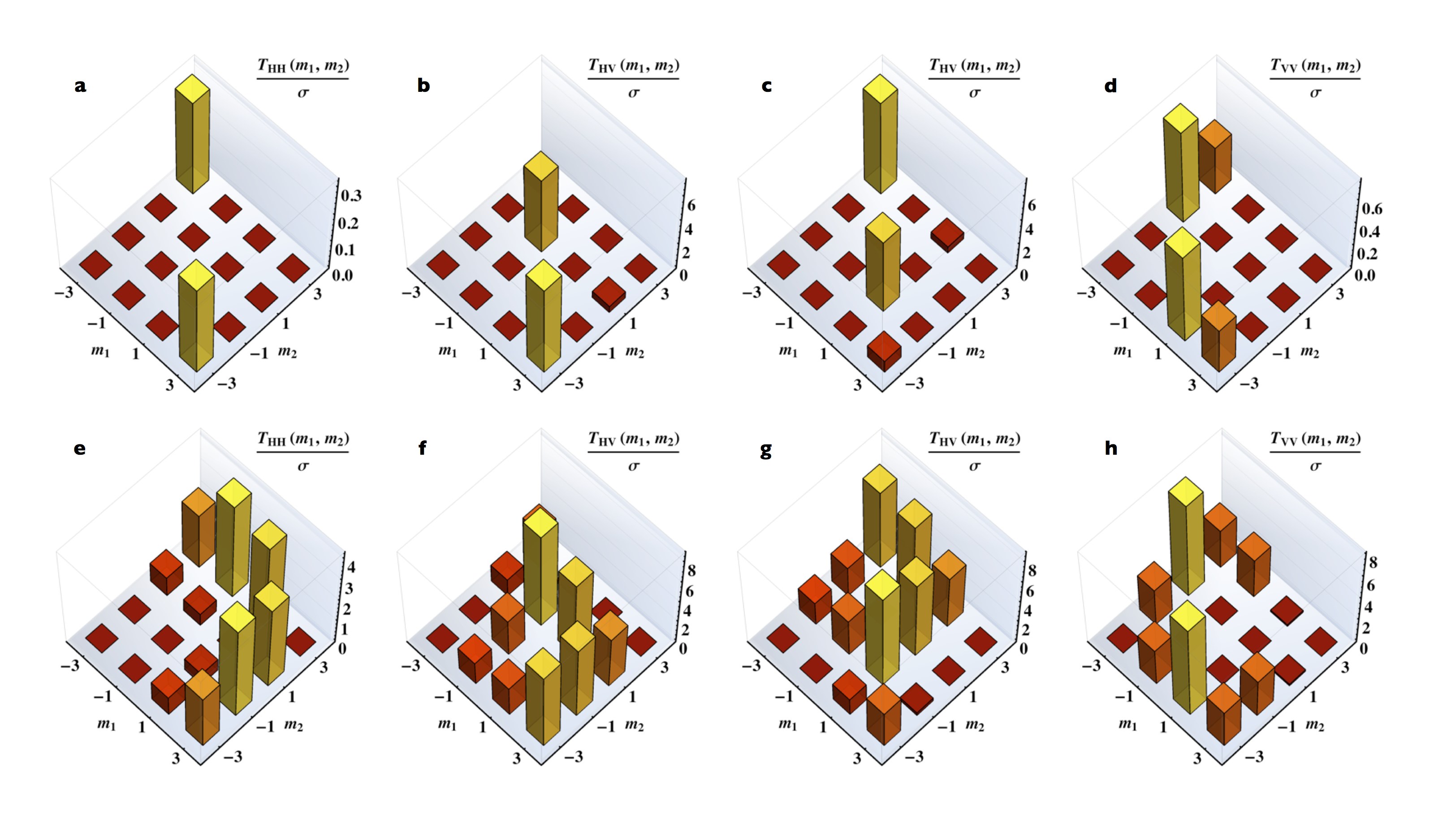}
\caption{Experimental violation of correlation inequalities for two photons which have completed the standard QW ($\delta=\pi$). The data are based on the coincidences after the final beam-splitter. (a)-(d) Violations of the inequalities given in Eq.\ (\protect\ref{classicalinequality}), constraining the correlations that would be obtained for two classical sources, incoherent to each other. Each panel refers to a different pair of measured polarizations for the two photons. These violations prove that our results can only be explained with quantum effects. (e)-(h) Violations of the inequalities given in Eq.\ (\protect\ref{photoninequality}), constraining the correlations obtained for two distinguishable photons. Again, each panel refers to a different pair of polarizations. These violations prove that our photons exhibit two-particle interferences. Only positive values of the $T_{p,q}$ are reported, while negative values which fulfil the inequality are omitted. All violations are given in units of Poissonian standard deviations $\sigma$, as determined from the coincidence counts. The color scale reflects the vertical scale.}
\label{fig:ineqstandard}
\end{figure*}

\begin{figure*}[t]
\centering
\includegraphics[width=18cm]{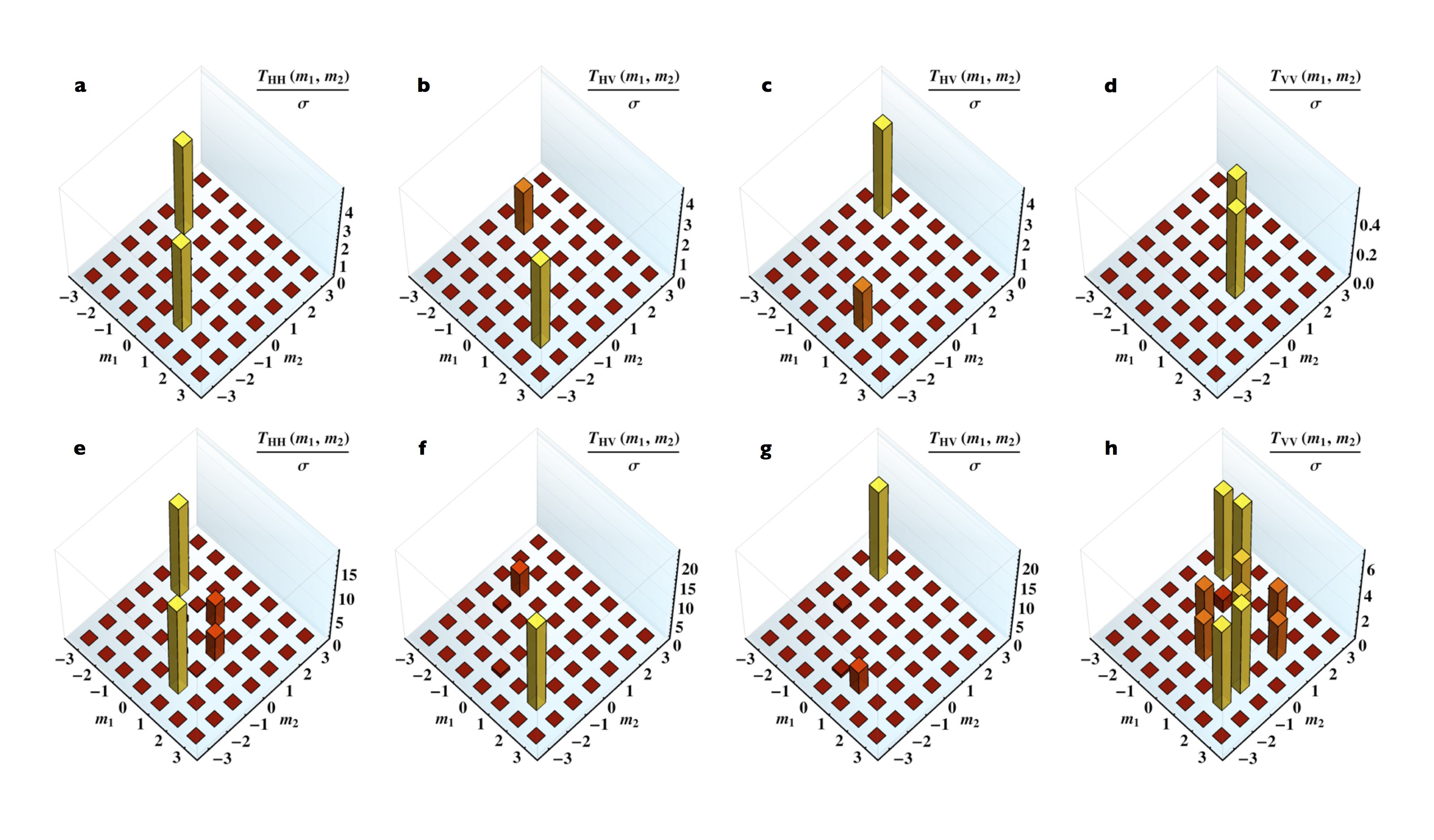}
\caption{Experimental violation of correlation inequalities for two photons which have completed the hybrid QW ($\delta=\pi/2$). The data are based on the coincidences after the final beam-splitter. (a)-(d) Violations of the inequalities given in Eq.\ (\protect\ref{classicalinequality}), constraining the correlations that would be obtained for two classical sources, incoherent to each other. Each panel refers to a different pair of measured polarizations for the two photons. These violations prove that our results can only be explained with quantum effects. (e)-(h) Violations of the inequalities given in Eq.\ (\protect\ref{photoninequality}), constraining the correlations obtained for two distinguishable photons. Again, each panel refers to a different pair of polarizations. These violations prove that our photons exhibit two-particle interferences. Only positive values of the $T_{p,q}$ are reported, while negative values which fulfil the inequality are omitted. All violations are given in units of Poissonian standard deviations $\sigma$, as determined from the coincidence counts. The color scale reflects the vertical scale.}
\label{fig:ineqhybrid}
\end{figure*}

Let us now discuss the inequalities constraining the measurable photon correlations in two specific reference cases. Our first reference case is that of two independent classical sources (or coherent quantum states with random relative phases) entering modes 1 and 2, in the place of single photons. The following inequality can be then proved to apply to the intensity correlations $\Gamma_{p,q}=\langle \hat{a}^{\dag}_p\hat{a}^{\dag}_q\hat{a}_p\hat{a}_q\rangle$, for any two given QW output modes $p$ and $q$ \cite{Brom09_PRL,Peru10_Sci}:
\begin{equation}
\frac{1}{3}\sqrt{\Gamma_{p,p}\Gamma_{q,q}}-\Gamma_{p,q}<0.
\end{equation}
In terms of two-photon detection probabilities $\bar{P}_{p,q}=(1+\delta_{p,q})\Gamma_{p,q}$, the same inequality reads
\begin{equation}
\frac{2}{3}\sqrt{\bar{P}_{p,p}\bar{P}_{q,q}}-\bar{P}_{p,q}<0,
\end{equation}
where $\bar{P}_{p,q}$ stands for the probability of having state $\ket{1_p,1_q}$, for $p\neq q$, or state $\ket{2_p}$, for $p=q$, after the QW but before the BS used to split the photons. After the BS, taking into account the photon-splitting probability, the inequality is rewritten as
\begin{equation}
T_{p,q}=\frac{1}{3}\sqrt{P_{p,p}P_{q,q}}-P_{p,q}<0,
\label{classicalinequality}
\end{equation}
where $P_{p,q}$ is now the probability of detecting in coincidence a photon in state $p$ at one (given) BS exit port and the other photon in state $q$ at the other BS exit port.

Our second reference case is that of two single but distinguishable photons entering states 1 and 2. In this case, it is easy to prove a second stronger inequality for the coincidence probabilities. Indeed, in this case one has
\begin{equation}
\bar{P}_{p,q}=|U_{1,p}U_{2,q}|^2+|U_{1,q}U_{2,p}|^2
\end{equation}
for $p\neq q$ and
\begin{equation}
\bar{P}_{p,p}=|U_{1,p}U_{2,p}|^2,
\end{equation}
where $\bar{P}_{p,q}$ now stands for the probability of having one of the two distinguishable photons in state $p$ and the other in $q$ after the QW, before the BS.
The mathematical identity $(|U_{1,p}U_{2,q}|-|U_{1,q}U_{2,p}|)^2>0$ leads directly to the following inequality:
\begin{equation}
2\sqrt{\bar{P}_{p,p}\bar{P}_{q,q}}-\bar{P}_{p,q}<0.
\end{equation}
After the BS, this in turn is equivalent to
\begin{equation}
T_{p,q}=\sqrt{P_{p,p}P_{q,q}}-P_{p,q}<0.
\label{photoninequality}
\end{equation}

The violation of the first inequality (\ref{classicalinequality}) from our coincidence data would prove that the photon correlations cannot be mimicked by intensity correlations of classical sources. Panels (a-d) in Figs.\ \ref{fig:ineqstandard} (standard QW) and \ref{fig:ineqhybrid} (hybrid QW) show the set of violations found in our two-photon experiments, in units of Poissonian standard deviations. In some cases, the experimental violations are larger than 5 standard deviations, proving that the measured correlations are quantum. As these inequalities are valid for any possible unitary propagation of the photons, they are also independent of all possible misalignments of our setup. Hence, the use of statistical standard deviations to assess the violation magnitude is well justified.

The violation of the second inequality (\ref{photoninequality}) from our data proves that the photon correlations are stronger than those allowed for two distinguishable photons, owing to the contribution of two-photon interferences. Although this is already demonstrated in some cases by the violation of the first inequality (as the violation of the first inequality logically implies the violation of the second one), this second inequality is stronger and should be therefore violated in a larger number of cases and with a larger statistical significance (although it requires assuming that there are two and only two photons at input, so that a classical source is excluded a priori). Panels (e-h) in Figs.\ \ref{fig:ineqstandard} (standard QW) and \ref{fig:ineqhybrid} (hybrid QW) show the observed violations. This time, certain measurements violate the inequality by as much as 15 standard deviations, thus proving that two-photon interferences play a very significant role in our experiment.

\end{document}